\documentclass[15pt a4paper]{article}

\usepackage{latexsym}
\usepackage{amsmath}
\usepackage{amssymb}
\usepackage{graphicx}
\graphicspath{{./}{./figs/}}

\begin{document}

\title{\bf A comprehensive weighted evolving network model\thanks{Accepted by Physica A, May 2004}}
\author{Chunguang Li$^1$\thanks{Corresponding author, Email:
cgli@uestc.edu.cn}, Guanrong Chen$^2$}

\date{\small $^1$Institute of Electronic Systems, School of Electronic
Engineering, \\University of Electronic Science and
Technology of China,\\ Chengdu, 610054, P. R. China.\\
$^2$Department of Electronic Engineering, City University of Hong
Kong, \\83 Tat Chee Avenue, Kowloon, Hong Kong, P. R. China.}
\maketitle

\begin{abstract}
Many social, technological, biological, and economical systems are
best described by weighted networks, whose properties and dynamics
depend not only on their structures but also on the connection
weights among their nodes. However, most existing research work on
complex network models are concentrated on network structures,
with connection weights among their nodes being either 1 or 0. In
this paper, we propose a new weighted evolving network model.
Numerical simulations indicate that this network model yields
three power-law distributions of the node degrees, connection
weights and node strengths. Particularly, some other properties of
the distributions, such as the droop-head and heavy-tail effects,
can also be reflected by this model.
\end{abstract}

Complex networks are currently being studied across many fields of
science and engineering [1], stimulated by the fact that many
systems in nature can be described by models of complex networks.
A complex network is a large set of interconnected nodes, in which
a node is a fundamental unit usually with specific dynamical or
information contents. Examples include the Internet, which is a
complex network of routers and computers connected by various
physical or wireless links; the World Wide Web, which is an
enormous virtual network of web sites connected by hyperlinks; and
various communication networks, food webs, biological neural
networks, electrical power grids, social and economic relations,
coauthorship and citation networks of scientists, cellular and
metabolic networks, etc. The ubiquity of various real and
artificial networks naturally motivates the current intensive
study of complex networks, on both theory and applications.

Many properties of complex networks have been reported in the
current literatures. Notably, it is found that many complex
networks show the small-world property [2], which implies that a
network has a high degree of clustering as in a regular network
and a small average distance between nodes as in a random network.
Another significant recent discovery is the observation that many
large-scale complex networks are scale-free. This means that the
degree distributions of these complex networks follow a power law
form $P(k)\sim k^{-\gamma}$ for large $k$, where $P(k)$ is the
probability that a node in the network is connected to $k$ other
nodes and $\gamma$ is a positive real number determined by the
given network. Since power laws are free of characteristic scale,
such networks are called ``scale-free networks'' [3]. The
scale-free nature of many real-world networks can be generated by
a mechanism of growing with preferential attachment [3].

In most growing network models, all the links are binary with
values being either 1 or 0. However, many real-world networks,
such as various scientific collaboration networks [4] and
ecosystems [5], display different interaction weights between
nodes. Recently, some weighted evolving network models are
proposed. In [6], a weighted network model was presented, in which
both the structure and the connection weights are driven by the
connectivity according to the preferential attachment rule. In
[7], four weighted evolving network models were proposed and
analyzed. In [8], a weighted evolving network model with
stochastic weight assignments was suggested. More recently,
additional information about the connection weights in real-world
complex networks has been found. In [9], for example, we found
that the connection weights of many real-world networks indeed
obey power-law distributions. In [10], it was also found that the
strengths of nodes in complex networks obey a power-law
distribution. Here, the strength $s_i$ of node $i$ is defined as
[10]
\begin{equation}
s_i=\sum_{j\in V(i)}w_{ij}
\end{equation}
where the sum runs over the set $V(i)$ of some specified neighbors
of node $i$. Thus, together with the power-law distribution of
degrees, there are three power-law distributions concerning the
structure and the connection in a complex network according to
statistical studies. The weighted network models considered in the
aforementioned references reflect only part of these three
power-law distributions.

It should also be noted that, most existing weighted network
models are driven entirely by the preferential attachment scheme
in structure and the preferential strengthening scheme in
connection weights. In real-world networks, other than these kinds
of ``preference", there are also ``randomness". For example, in
scientific collaboration networks, if a new comer is a student to
enter the field, he/she usually collaborates with his/her advisor
other than a popular author unknown to him/her in this field. In
this case, the nodes that new nodes attached to are likely being
uniformly distributed. This can be described by {\it random
attachment}. Moreover, if two individuals collaborate successfully
many times, they are likely to keep on collaborating and even
increase their joint work. This is the effect of {\it preferential
strengthening}. But, sometimes, this preferential strengthening is
disobedient. For example, if an individual moved to a new
university, he/she is most likely to begin to collaborate with new
colleagues in the same field, and stop or reduce the collaboration
with his/her original co-authors. This can be described by {\it
random strengthening}. In this paper, we propose a new weighted
evolving network model, in which we consider both the preferential
effects and the random properties of complex networks. This
network model displays all the three power-law distributions
mentioned above. Besides, some other distribution properties of
networks, such as the droop-head (in the foreside of the power-law
distribution, data are usually located below the declining line
with a slope $-\gamma$) and heavy-tail (in the endside of the
power-law distribution, data are usually widely spread out)
effects can also be reflected by this new network model. Existing
network models generally cannot reflect these properties.
Recently, in [11], the authors explained the ``droop-head" shape
of $P(k)$ distributions in complex networks by using nonextensive
entropy approach. But it seems that no explicit connection exists
between their approach and our network model.

For simplicity, we only consider undirected network models.
Directed weighted evolving network model will be studied
separately. The proposed model is defined by the following scheme:
\medskip

{\bf Step 1.} Start from a small number $m_0$ ($m_0>1$) of fully
connected nodes.

{\bf Step 2.} At each time step, pick a preferred probability
$\alpha\in[0,1]$ and, with a uniform distribution, randomly
generate a real number $s\in[0,1]$. If $s\le\alpha$, then a new
node of strength 1 is added with probability $\alpha$, and this
new node is connected to an existing node which is selected from
among all the existing nodes. This existing node, to which the new
node is connected, is selected according to the rule specified at
Step 3 below. But if $s>\alpha$, then no new node will be added.
Instead, with probability $1-\alpha$, a new connection with weight
1 is added between two existing nodes, where these two existing
nodes are selected according to the rule specified at Step 4
below.

{\bf Step 3.} Pick a preferred probability $\beta\in[0,1]$ and,
with a uniform distribution, randomly generate a real number
$r\in[0,1]$. If $r\le\beta$, then with probability $\beta$ an
existing node is selected with the following probability:
\begin{equation}
\Pi(k_i)=\frac{k_i}{\sum_j k_j}
\end{equation}
where $k_i$ is the degree of node $i$ ({\it preferential
attachment}); but if $r>\beta$, then with probability $1-\beta$ an
existing node is selected randomly ({\it random attachment}).
(Note: this selected node is the ``existing node that the new
added node is connected to'' discussed in Step 2 above.)

{\bf Step 4.} Pick a preferred probability $\eta\in[0,1]$ and,
with a uniform distribution, randomly generate a real number
$\xi\in[0,1]$. If $\xi\le\eta$, then with probability $\eta$ two
existing nodes are chosen from among all existing ones, with the
following probability:
\begin{equation}
\Pi(i,j)=\frac{w_{ij}}{\sum_{k,l}w_{kl}}
\end{equation}
where $w_{ij}\,(=w_{ji})$ is the connection weight value between
nodes $i$ and $j$ ({\it preferential strengthening}); but if
$\xi>\eta$, then with probability $1-\eta$ two existing nodes are
chosen randomly ({\it random strengthening}).
\medskip

After $N$ time steps, this scheme generates a network with
$m_0+N\alpha$ nodes in the sense of mathematical expectation and
the sum of total connection weights is
$\frac{1}{2}\sum_{i,j}w_{ij}= N+\frac{1}{2}m_0!$. Parameter
$\alpha$ controls the ratio of growing and strengthening; $\beta$
and $\eta$ control the ratios between ``preference" and
``randomness". In reality, different networks have different
values of these parameters, specified by the nature of the given
network.

\begin{figure}[htb]
\centering
\includegraphics[width=7cm]{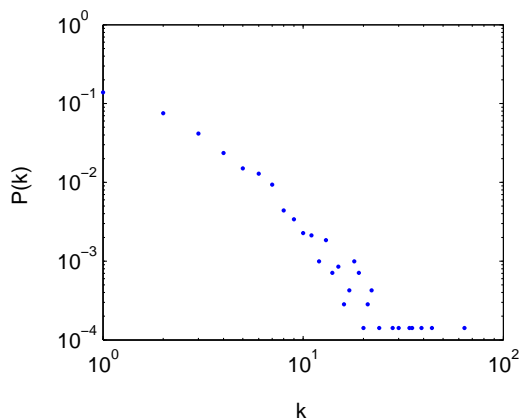}
\caption{Probability distribution $P(k)$ of degree $k$ ($m_0=3,
N=8000, \alpha=0.3, \beta=\eta=0.8$).}
\end{figure}
\begin{figure}[htb]
\centering
\includegraphics[width=7cm]{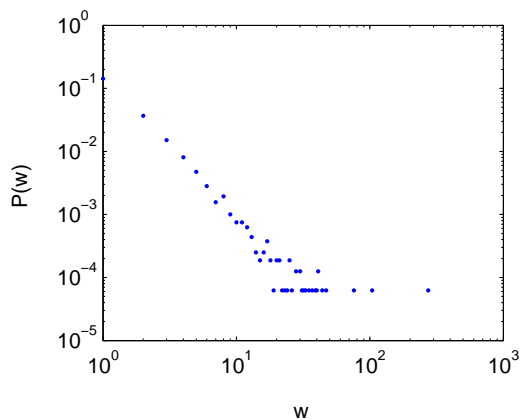}
\caption{Probability distribution $P(w)$ of the connection weight
$w$ ($m_0=3, N=8000, \alpha=0.3, \beta=\eta=0.8$).}
\end{figure}
\begin{figure}[htb]
\centering
\includegraphics[width=7cm]{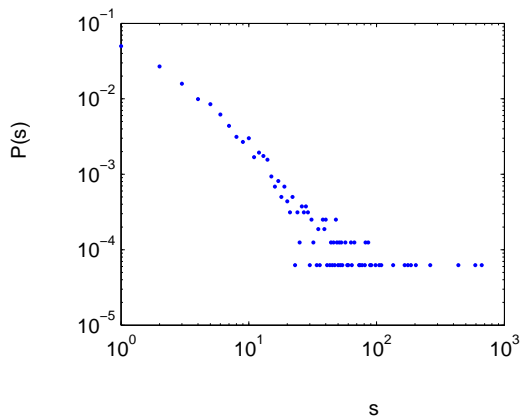}
\caption{Probability distribution $P(s)$ of the node strength $s$
($m_0=3, N=8000, \alpha=0.3, \beta=\eta=0.8$).}
\end{figure}
\begin{figure}[htb]
\centering
\includegraphics[width=7cm]{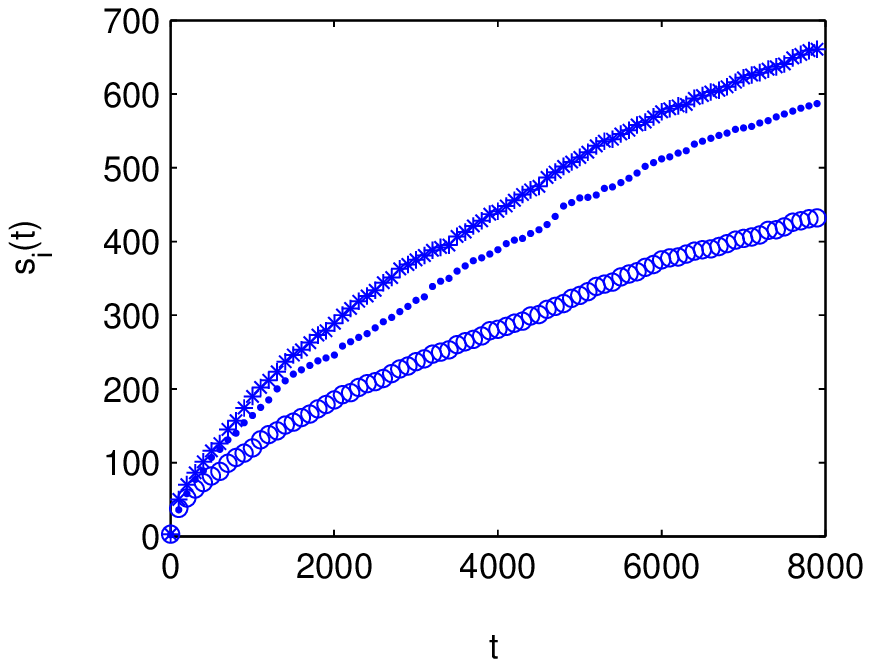}
\caption{Time evolution of the strengths of three initial nodes.}
\end{figure}

We performed numerical simulations with different parameter values
of $\alpha, \beta,\eta$ and $N$. Here, we show the simulation
results with $m_0=3$ and probabilities $\alpha=0.3,
\beta=\eta=0.8$. First, we study the degree distribution $P(k)$,
which is defined as the probability that a randomly selected node
has degree $k$ (a node has $k$ connections). In Fig.1, we show
that the degree distribution $P(k)$ obeys a power-law distribution
$k^{-\gamma}$ for a large range of $k$ with exponent
$\gamma\approx 2.8$, which has a droop-head and a heavy-tail. This
is consistent with the statistical results of many real-world
networks [3]. In Fig.2, we show the connection weight distribution
$P(w)$, which is defined as the probability that a randomly
selected link between two nodes has a weight value $w$. As can be
seen from Fig.2, this distribution also behaves in a power-law
form, with exponent $\gamma\approx 3$, which is also similar to
many real-world data [9]. In Fig.3, we show the node strength
distribution $P(s)$, which is defined as the probability that a
randomly selected node has a strength value $s$. This distribution
also obeys a power-law form for a large range of strength values,
with exponent $\gamma\approx 2.1$. It also has a droop-head and a
heavy-tail, which again coincides with the statistical results of
many real-world networks shown in [10]. In Fig.4, we show the time
evolution of the strengths of three initial nodes. As can be seen,
the increasing speeds of the strength curves decrease gradually as
$t$ goes on. In real-world networks, this is also the case due to
limited resources and competition etc.

The above simulation results show that our new network model can
indeed mimics many real-world complex networks comprehensively.
For networks with non-integer connection weight values, such as
neural networks, we can also mimic them by the proposed model,
using proper non-integer constants as the weight values in each
step of the evolving process.

Recently, we noticed that in [12] the authors gave a complex
network model with both preferential and random attachments.
However the network model considered in [12] is not a weighted
network model, and the evolving scheme is fundamentally different
from ours.

In summary, to reflect the growing and strengthening dynamics with
``preference" and ``randomness", we have proposed a comprehensive
weighted evolving network model. Simulation results show that this
model can displays all the three power-law distributions regarding
the network structure and connection strengths observed in
statistical studies of many real-world networks. Also, the
droop-head and heavy-tail properties of these distributions, which
are observed in many real-world networks, can be reflected by this
new network model. In comparison, existing evolving network models
generally cannot describe all these features together. Thus, in
this paper, we have provided an effective model to mimic many
real-world weighted evolving networks.

This research was supported by the Hong Kong Research Grant
Council under the CERG grant CityU 1115/03E, the National Natural
Science Foundation of China under Grant 60271019, and the Youth
Science and Technology Foundation of UESTC under Grant YF020207.

\end{document}